\documentclass{ws-ijmpa}

\RequirePackage{xspace}

\def\epem {\ensuremath{e^+e^-}\xspace}

\newcommand{\tev}{\ensuremath{\mathrm{\,Te\kern -0.1em V}}\xspace}
\newcommand{\gev}{\ensuremath{\mathrm{\,Ge\kern -0.1em V}}\xspace}
\newcommand{\mev}{\ensuremath{\mathrm{\,Me\kern -0.1em V}}\xspace}
\newcommand{\kev}{\ensuremath{\mathrm{\,ke\kern -0.1em V}}\xspace}
\newcommand{\ev}{\ensuremath{\mathrm{\,e\kern -0.1em V}}\xspace}
\newcommand{\gevc}{\ensuremath{{\mathrm{\,Ge\kern -0.1em V\!/}c}}\xspace}
\newcommand{\mevc}{\ensuremath{{\mathrm{\,Me\kern -0.1em V\!/}c}}\xspace}
\newcommand{\gevcc}{\ensuremath{{\mathrm{\,Ge\kern -0.1em V\!/}c^2}}\xspace}
\newcommand{\mevcc}{\ensuremath{{\mathrm{\,Me\kern -0.1em V\!/}c^2}}\xspace}
\def\micron{\ensuremath{\mu{\rm \,m}}\xspace}

\def\nb         {\ensuremath{{\rm \,nb}}\xspace}

\def\invfb   {\ensuremath{\mbox{\,fb}^{-1}}\xspace}

\newcommand{\dedx}{\ensuremath{\mathrm{d}\hspace{-0.1em}E/\mathrm{d}x}\xspace}
\def\mum  {\ensuremath{{\,\mu\rm m}}\xspace}

\def\Bbar    {\kern 0.18em\bar{\kern -0.18em B}{}\xspace}
\def\BBbar      {\ensuremath{B\Bbar}\xspace}
\def\Bfactory {$B$-Factory\xspace}
\def\Bfactories {$B$-Factories\xspace}
\def\invcminvs{\ensuremath{\rm cm^{-2} \, \rm s^{-1}}\xspace}
\def\DDbar {\ensuremath{D^{(*)} \bar {D}^{(*)}}\xspace}
\begin{document}

\markboth{F.~Renga}
{Signatures of Exotic Hadrons}

%
\catchline{}{}{}{}{}
%

\title{SIGNATURES OF EXOTIC HADRONS}

\author{FRANCESCO RENGA}

\address{Paul Scherrer Institut PSI, CH-5235 Villigen, Switzerland\\ 
and\\
INFN Sezione di Roma\\
francesco.renga@roma1.infn.it}

\maketitle

\begin{history}
\received{Day Month Year}
\revised{Day Month Year}
\end{history}

\begin{abstract}
Hadron spectroscopy represented in the past a major tool for understanding the fundamental symmetries of strong forces.
More recently, the interest on this topic has been revitalized by the discovery of new quarkonium-like resonances, that do not fit in the 
standard picture and whose understanding could improve our mastery of quantum chromodynamics. 
I review here the experimental signatures of these exotic hadrons, at present and future \epem and hadron collider experiments.

\end{abstract}

\ccode{PACS numbers: 14.40.Rt, 14.40.Pq, 14.40.Be}

\section{Introduction}

Even before the general acceptance and the experimental proof of the quark hypothesis, hadron spectroscopy provided a formidable insight into the
fundamental properties of both strong and weak interactions. The identification of the pion system as an isospin triplet and the \emph{Eightfold way} 
extension of this formalism led to the first formulation of the constituent quark model;\cite{eightfold} the study of the $\Omega^-$ and $\Delta^{++}$ 
baryons provided one of the first indications of an additional quantum number for quarks: the color;\cite{color} later on, the discovery of the $J/\psi$ 
resonance\cite{psi,J} and its spectroscopic properties provided the first evidence for the charm quark, while the charmonium spectrum still provide to date a 
solid testing ground for effective QCD theories. These are just a few examples of the wide impact of hadron spectroscopy on the development of particle
physics.

On the other hand, until a few years ago the experimental situation in this field appeared quite stable and well understood. Almost all observed resonances 
could be clearly identified as two- or three-quark states, with masses, widths and quantum numbers in reasonable agreement with the expectations 
from QCD. This was true, in particular, for the quarkonium states (for which a memorandum of the naming conventions is provided in~\ref{app:naming}). 
The only exceptions were a few scalar mesons, the $a_0$'s and $f_0$'s resonances, whose properties will be discussed later.

The situation changed dramatically in the last ten years. In 2003 the Belle collaboration claimed the observation of a new resonance around 
3.872 \gevcc, decaying into $J/\psi \,\pi^+ \pi^-$,\cite{X3872_belle} named since then $X(3872)$. The result was soon confirmed by the BaBar 
collaboration.\cite{X3872_babar} The $J/\psi \, \pi^+ \pi^-$ decay channel suggests a charmonium-like structure, but this state does not fit in 
the standard charmonium model: its mass ($3872.2 \pm 0.4$ \mevcc) is far from any predicted charmonium state and its width ($3.0^{+1.9}_{-1.4} \pm 0.9$ \mev) 
is far too small for a standard charmonium lying above the open charm threshold. In the subsequent years, a profusion of similar states
was found. 

This review is devoted to the experimental signatures of these exotic states. In Sec.~\ref{sec:theo} I will provide a short theoretical introduction 
to the most popular models proposed to explain the properties of the new resonances. The experimental techniques behind their discovery and the assessment 
of their properties are described in Sec.~\ref{sec:exp}. Finally, a review of the observed states, with possible interpretations, is provided in Sec.~\ref{sec:states}.

\section{Theoretical overview}
\label{sec:theo}

Several reviews have been already published to discuss the theoretical aspects behind the study of the exotic resonances (see for instance Ref.~\refcite{review} and
Ref.~\refcite{reviewNC}). Hence, I give here only a short theoretical overview, before concentrating on the experimental aspects of this field.

The theoretical foundations of the standard hadron spectroscopy came from these two features of QCD:
\begin{enumerate}
\item Only color-singlet states can exist as observable particles (\emph{confinement});
\item Only some combinations of color states produce an attractive potential, leading to a bound state.  
\end{enumerate}
In particular, the combination of a color and an anti-color triplets, $\mathbf{3}_C \otimes \mathbf{\bar 3}_C$ can produce 
a singlet and an octet, $\mathbf{1}_C \oplus \mathbf{\bar 8}_C$, characterized by an attractive and a repulsive potential, respectively. Hence,
a quark and an anti-quark can compose a a bound state $q \bar q'$ in a singlet state, which can exist as an observable particle: these are the standard
mesons. Similarly, $\mathbf{3}_C \otimes \mathbf{3}_C \otimes \mathbf{3}_C = \mathbf{10}_C \oplus \mathbf{8}_C \oplus \mathbf{8}_C \oplus \mathbf{1}_C$, where
again the singlet has an attractive potential and bound states $qqq$ can be built: these are the standard baryons. Anti-mesons and anti-baryons are similarly 
obtained.

Anyway, other bound states are predicted by QCD, apart from mesons and baryons. In particular, the combination of two color triplets,
$\mathbf{3}_C \otimes \mathbf{3}_C$, produces a sextet with repulsive potential, but also a color triplet with attractive potential. When such
a state is obtained with two quarks, it is called a \emph{diquark}, $[qq']$. The color triplet obtained in this way can be combined with an anti-color 
triplet and, according to the rules above, it can compose a color singlet with attractive potential. In particular, a diquark and an anti-diquark
can compose a color singlet in a bound state: it is called a \emph{tetraquark}. A further possibility is to combine a gluon (that is a color octet) with the
$\mathbf{\bar 8}_C$ of a $q \bar q'$ pair. Also in this case a color singlet bound state can be obtained: a \emph{hybrid} $gq \bar q'$. In addition,
an attractive potential can be also obtained between color singlets: it allows to build bound states of mesons (i.e. \emph{meson molecules}). Finally, more
complex combinations like pentaquarks or gluon-gluon states (\emph{glueballs}) are also allowed.

\begin{figure}[pb]
\centerline{\psfig{file=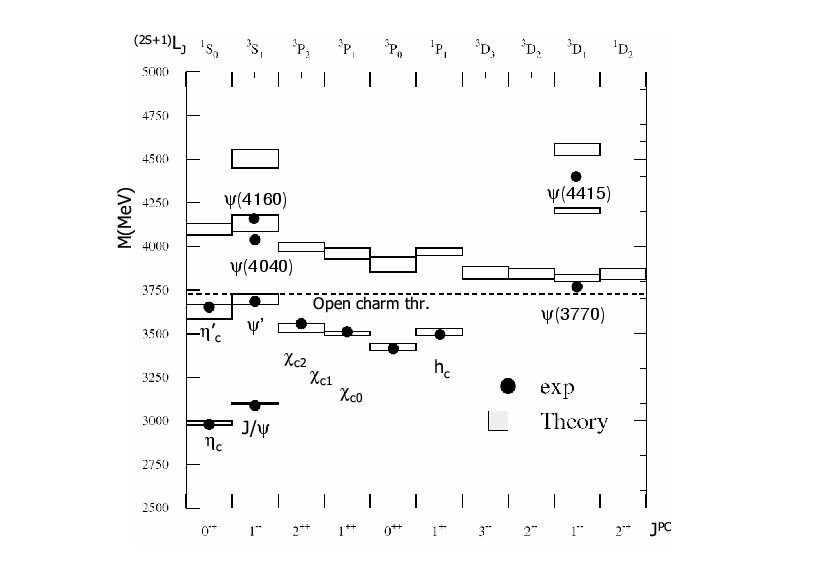,width=9cm}}
\vspace*{8pt}
\caption{Summary of expected and observed regular charmonia. \label{fig:charmonia}}
\end{figure}

Among regular mesons, heavy quarkonia $c \bar c$ and $b \bar b$ are of particular interest. Due to the large mass of the charm and bottom quarks,
these objects can be treated in non-relativistic QCD approximations, that yielded since the late 1970 quite precise estimates of their masses and widths.\cite{eichten}
Most often, in these models, an effective quark-anti-quark potential is defined, the most famous being the Cornell potential $V(r) = k/r + a \cdot r$, reproducing
the most relevant feature of QCD (strong attraction at small distances and color confinement at large distances); then, the parameters of the potential are fixed
looking at the lowest mass quarkonia and are used to predict the masses of the other states. More recently, these results have been confirmed by lattice 
calculations (see the lattice section in Ref.~\refcite{review}). A summary of the expected and observed regular charmonia can be found in Fig.~\ref{fig:charmonia},
where the expectations are taken from Ref.~\refcite{HQP_review}.
Hence, it is quite easy to determine the exotic nature of charmonium-like and bottomonium-like states: if a new state emerges, whose decay modes 
suggest a quarkonium-like content, but which escapes any mass and width predictions of regular quarkonia, it can be considered a good candidate for an 
exotic resonance. In this respect, I already made, in the introduction, the example of the $X(3872)$ resonance. The dominance of some decay modes over 
the ones that are expected to be favored for regular charmonia is another important indication.

The difficult part of the game is to discriminate among different hypotheses for the composition of such exotic states. In the next section I will illustrate the
most important experimental observables that can help in this task.

\subsection{Tetraquarks, hybrids and molecules: experimental discriminants}
\label{sec:discriminants}

Beside the different mass predictions that are obtained for tetraquarks, hybrids and molecules, other discriminant observables exist, that can be used to
determine the nature of the exotic states.

At first, the allowed flavor and quantum number combinations are different in the three cases. In particular, in the tetraquark hypothesis a very large number
of new states is possible, and they would be organized in flavor multiplets, whose members would have a very similar mass. The $J^{PC}$ quantum numbers
of such states can assume non-standard values. Moreover, among the possible combinations, there are several that produce charged resonances, that 
cannot be otherwise composed with less than four quarks; they would be associated to neutral partners of similar mass, producing a clear signature of a tetraquark 
multiplet. Finally, tetraquarks show large decay branching ratios (BR) into modes other than \DDbar ($X \to J/\psi \pi \pi$ is a typical example) and 
a smaller total width with respect to the regular charmonia above the open charm threshold.

The variety of hybrid states is expected to be much smaller, but again non-standard quantum numbers can emerge, namely $J^{PC} = 1^{-+}$. Unlike the
tetraquarks, these resonances are expected to decay almost exclusively into \DDbar.

This decay mode is also the most natural for \DDbar meson molecules, that anyway can still have small decay widths even above the open
charm threshold. The most natural width for a $D \bar {D}^{*}$ molecule is actually of the same order of the $D^{*} \to D \pi$ one 
($\sim$70~\kev),\cite{reviewNC} much lower than what is expected even for a tetraquark. This result provides an interesting way to discriminate 
between the two hypotheses, and I will apply it to the interpretation of the $X(3872)$. The molecules are also expected to appear not so far from some 
\DDbar threshold, and only a few states are possible, with well defined quantum numbers, derived by the combination of the meson quantum numbers.

\section{Experimental methods}
\label{sec:exp}

\subsection{Machines and experiments}

Searches for exotic hadrons can be performed both at \epem and hadron colliders. In fact, most of the exotic states
observed in the last few years have been found by the \Bfactory experiments BaBar, Belle and CLEO, but very important information
and new observations also came from the $\tau$-charm factory BEPC and from the hadron colliders, with Tevatron experiments that are on the route 
to exploit their full statistics and LHC that already started to produce interesting results. In a longer time scale, the devised \emph{Super Flavor Factories}
should provide a lot of additional information. A short introduction to these experiments is given in this section.

\subsubsection{\Bfactories}

A \Bfactory is an \epem machine running at a center of mass (CM) energy of about 10.58~\gev, corresponding to the mass of the $\Upsilon(4S)$ resonance, which
decays almost exclusively to \BBbar pairs, with a cross section of about 1.1~\nb. Three facilities have been operated at this energy in the last two 
decades: CESR at Cornell (USA), PEP-II at SLAC (USA) and KEK-B at KEK (Japan). The latter two are \emph{asymmetric} \Bfactories, where the two colliding 
beams are different in energy and the \BBbar pair is produced with a Lorentz boost in the laboratory frame. Peak luminosities of $1.2 \times 10^{33}$, 
$12 \times 10^{33}$ and $21 \times 10^{33}$ \invcminvs were reached by the three machines, respectively.

The CLEO, BaBar and Belle experiments have been operated at CESR, PEP-II and KEKB, respectively. In their final configuration, 
all experiments are provided of a silicon strips vertex detector and a drift chamber in a 1.5 T magnetic field, for internal tracking. 
A vertex resolution of $\sim$100 \micron is obtained, with a typical momentum resolution of 1\% for charged tracks at $\sim$4~\gevc. 
The \dedx is also measured for particle identification (PID). Photon and electron energies are measured by electromagnetic calorimeters built 
of CsI(Tl) crystals, providing an energy resolution of the order of 4\% at $\sim$1~\gevc. Kaon-pion separation above 700 keV is enhanced 
by dedicated PID systems: a ring imaging \v{C}erenkov detector (RICH) at CLEO, a detector of internally reflected \v{C}erenkov light (DIRC) at BaBar and a 
time-of-flight (TOF) system at Belle. Finally, streamer tubes (in CLEO) or resistive plate chambers (in BaBar and Belle), embedded in the iron of the
magnetic flux return, were used for muon identification and tracking.

The three experiments collected most of their data at the $\Upsilon(4S)$ peak. Furthermore, the beam energy was periodically lowered to collect events below the 
$B \bar B$ threshold, for an amount of about 10\% of the on-peak integrated luminosity, in order to study the background contributions from continuum 
$\epem \to q \bar q$ events. Finally, BaBar and Belle also collected relevant amounts of data at the $\Upsilon(2S)$, $\Upsilon(3S)$ and above the 
$\Upsilon(4S)$ (up to the energy of the candidate $\Upsilon(6S)$ peak), mostly for spectroscopic studies. CESR energy was instead lowered to convert 
it into a $\tau$-charm factory (see below). The integrated luminosities collected at the different energies by BaBar and Belle are reported in Table~\ref{tab:lumi}.

\begin{table}[ph]
\tbl{Collected luminosities (in \invfb) at the different CM energies by the BaBar and Belle Collaborations.}
{\begin{tabular}{@{}ccc@{}} \toprule
 & BaBar & Belle \\
\colrule
$\Upsilon(4S)$  & 433 & 711 \\ 
off-peak  &  54 & 100 \\
$\Upsilon(1S)$  & - & 6 \\
$\Upsilon(2S)$  & 14 & 25 \\
$\Upsilon(3S)$  & 30 & 3 \\
above $\Upsilon(4S)$ & 4 (energy scan) & 121 (at the $\Upsilon(5S)$) \\
\botrule
\end{tabular} \label{tab:lumi}}
\end{table}

In the future, two projects aim at collecting up to 100 times more statistics than the recent \Bfactories: the KEK-B upgrade
SuperKEK-B and the SuperB project in Rome (Italy). The former should reach a peak luminosity of $8 \times 10^{35}$~\invcminvs, and the detector will be an 
upgrade of the Belle detector. The latter aim at $10^{36}$~\invcminvs, exploiting the innovative \emph{crab waist} technique to squeeze the beams and reach 
very high luminosities with a relatively low beam current.\cite{superB} The detector design is based in this case on the BaBar detector, with reuse of several components.

\subsubsection{$\tau$-charm Factories}

A $\tau$-charm factory is an \epem collider running at a CM energy around the $\tau$ and charm production threshold, $\sim 3.5$~\gev.
Two machines where recently operated as $\tau$-charm factories: BEPC at IHEP (China) and CESR. They reached a peak luminosity 
of $12.6 \times 10^{30}$ and $76 \times 10^{30}$~\invcminvs, respectively.

Three version of the BES detector were adopted at BEPC. The design of the BES-II version involved a straw tube system and a drift chamber for internal tracking, a
TOF device for PID and a sampling electromagnetic calorimeter, composed by streamer tubes and lead absorbers.
Proportional tubes were used for the muon detector. A CsI(Tl) calorimeter, an upgraded TOF and an RPC-based muon detector were adopted in the BES-III version.

The CLEO detector was instead adapted to the lower beam energies, by replacing the silicon vertex detector with an inner drift chamber and reducing the
magnetic field to 1 T. This version of the experiment is known as CLEO-$c$.

The SuperB collider is designed to be also operated at the $\tau$-charm threshold.

\subsubsection{Hadron colliders and fixed-target experiments}

Hadron colliders are also a good place to study quarkonium spectroscopy and look for exotic resonances. The Tevatron $p \bar p$ collider operated
from 1987 to 2011. It started taking data at $\sqrt{s} = 1.76$~\tev in 2002, and two experiments, CDF-II and D0, have been operated there. 
Since 2009, the Large Hadron Collider also entered the game, running at $\sqrt{s} = 7$~\tev with a peak luminosity that reaches $2 \times 10^{33}$~\invcminvs.

The typical design of the detectors hosted at Tevatron and LHC includes an inner tracker in a solenoidal magnetic field, composed by silicon detectors 
(strips or pixels), an electromagnetic calorimeter, a hadronic calorimeter and a muon detector. At CDF-II and D0, the tracking is completed by a drift chamber
and a scintillating fiber detector, respectively. At ATLAS, a transition radiation tracker is also used. An impact point resolution of 80\mum is typically reached, with a
transverse momentum resolution around 10\% at 1~\gevc. With the only exception of CMS, electromagnetic calorimetry is performed by means of sampling devices. 
Typical resolutions are around 5\% at 1~\gev. Sampling devices are also used for hadron calorimetry. In this case, typical resolutions of 60 to 80\% are achieved 
at 1~\gev. Finally, the muon system is typically composed by tracking devices (gas- or scintillator-based) in the flux return of the solenoid field (at CDF and CMS) 
or in a dedicated toroidal field (at D0 and ATLAS).

Some special mentions are needed for the LHCb experiment. Being devoted to the study of bottom hadrons, the detector is arranged along the beam 
axis, in order to exploit the large bottom quark production cross section at small angles. Moreover, dedicated PID detectors are adopted for the reduction 
of the typical backgrounds affecting the flavor physics analysis (two RICH detectors, with gaseous and aerogel radiators).




Being operated with a peak luminosity up to of $4 \times 10^{32}$~\invcminvs, Tevatron provided since 2002 $\sim$12~\invfb of integrated luminosity.
Instead, an integrated luminosity of $\sim$1.6~\invfb has been provided as to August 2011 to the CMS, ATLAS and LCHb experiments by the LHC.

For what concern the fixed-target experiments, hadron spectroscopy and the search for exotica are among the main purposes of COMPASS. This experiment
is operated at the CERN SPS with high intensity hadron and muon beams, up to 260~\gev. The detector is composed of a beam spectrometer, 
for the measurement of the momentum of the incoming particle, equipped with silicon microstrips and scintillating fibers; a large-angle and 
a small-angle spectrometer for the interaction products, composed of gaseous detectors and scintillating fibers; a RICH detector for PID; two 
hadron calorimeters with iron absorbers and scintillating detectors; a lead glass electromagnetic calorimeter; two sets of tracking stations 
(equipped with multi-wire proportional chambers) and hadron absorbers (iron and concrete) for muon identification. The experiment is collecting data since 2002.

\subsection{Analysis Techniques}

In this section I will present the analysis techniques typically used in the searches for exotic hadrons. Most of them
are common to the searches for regular quarkonia, but such an overview is nonetheless important to better understand the
experimental issues behind the observation of the new states.

\subsubsection{Classification by production mechanism}

\begin{figure}[pb]
\vspace*{-1cm}
\centerline{\psfig{file=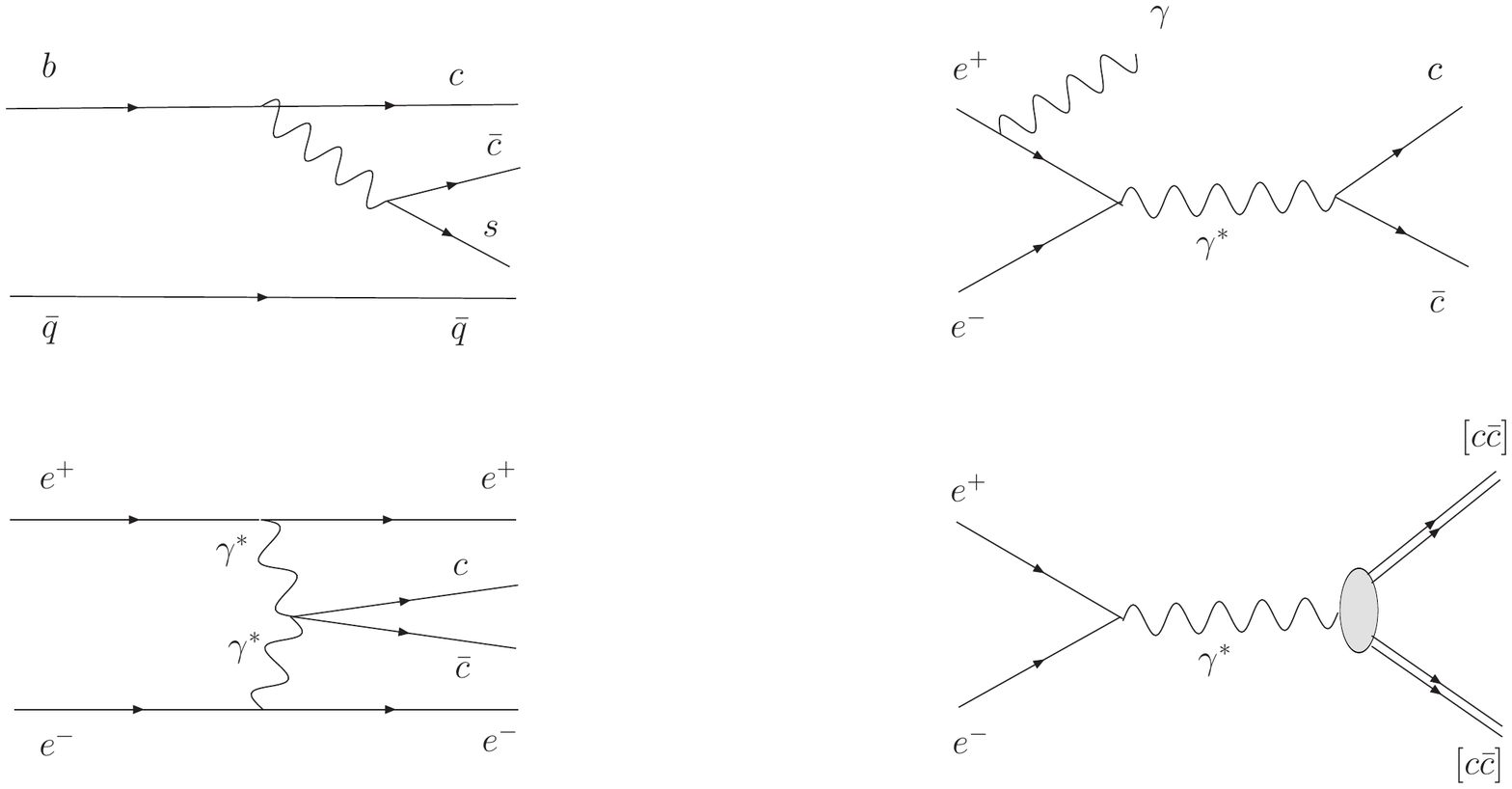,width=15cm}}
\vspace*{-3cm}
\caption{Production mechanisms for regular quarkonia and exotics at \epem colliders. From top left to bottom right: $B \to K h$, $s$-channel production with ISR,
two photon fusion and double quarkonium production. \label{fig:production}}
\end{figure}

The analysis technique adopted in the search for regular or exotic hadrons primarily depends on the production mechanism that
one wants to exploit. Four production mechanisms take place at \epem colliders (See Fig.~\ref{fig:production}):
\begin{enumerate}
\item the production in the decay of a $B$ meson, e.g. $B \to K h$;
\item the $s$-channel production $\epem \to \gamma^* \to h$, possibly with an initial state radiation (ISR), $\epem \to \gamma_{ISR} h$, where the hadron $h$, 
coming from the intermediate photon, has to carry $J^{PC} = 1^{--}$;
\item the two photon fusion production, $\epem \to \epem \gamma^* \gamma^* \to \epem h$;
\item the double quarkonium production, e.g. $\epem \to (c \bar c) (c \bar c)$.
\end{enumerate}

If the hadron is produced in the decay of a $B$ meson, the best strategy is to look at some exclusive channel, e.g. $B \to K J/\psi \pi \pi$. In this case, the 
$B$ signal is firstly separated from the continuum $\epem \to q \bar q$ ($q = u, d, s, c$) background by looking at two discriminating variables: the \emph{beam-energy substituted}
mass $m_{ES}$ and the \emph{missing energy} $\Delta E$:
\begin{eqnarray}
m_{ES} &=& \sqrt{(E_b^*)^2 - |\mathbf{p_B^*}|^2} \\
\Delta E &=& E_B^* - E_b^*
\end{eqnarray}
being $E_b^*$ the beam energy and $(E_B^*,\mathbf{p_B^*})$ the 4-momentum of the $B$ meson, both in the CM frame. For real $B$ decays, these two 
variables are expected to be 
centered at the $B$ mass and at zero, respectively. Once the $B$ meson is identified, peaks are searched for in the invariant mass distribution of the 
hadronic system, $J/\psi \pi \pi$ in our example. Notice that, in this and similar cases, once the $J/\psi$ has been identified in the $\ell^+\ell^-$ 
channel by an invariant mass cut, it is useful to constraint the lepton 4-momenta so that the reconstructed $J/\psi$ mass coincide with the nominal one, 
because it improves the mass resolution for the hadronic system. An alternative strategy, exploited by BaBar for the $B \to K h$ process,\cite{babarKrecoil} 
makes use of the recoil technique: one of the two $B$ mesons ($B_{reco}$) produced in the decay of the $\Upsilon(4S)$ is fully reconstructed in a hadronic final state 
and identified by means of an $m_{ES}$ selection; then, a kaon is searched for in the rest of the event. The full reconstruction of the $B_{reco}$, combined with the beam
energies, allows to estimate the 4-momentum of the other $B$ ($B_{sig}$), and the kaon momentum can be boosted in its rest frame. 
A peak in the distribution of the boosted momentum is an indication of a two-body decay $B_{sig} \to K h$, so that resonances can be searched for in a fully
inclusive way (i.e. with no assumption on their decay modes) and $BR(B \to K h)$ can be measured. At the present \Bfactories, this kind of
measurements turns out to be statistically limited. Anyway, at a Super \Bfactory, it could become a standard technique, taking advantage of the strong suppression
of the continuum background obtained by fully reconstructing the $B_{reco}$.

For the ISR production, the analysis technique depends on the difference between the $h$ mass and the CM energy. For large differences, the photon energy is 
large and the corresponding absolute resolution is also large. Hence, it is not convenient to exploit the measurement of the photon energy. Instead, one can look for
a peak in the invariant mass distribution of the hadronic system, possibly looking at some exclusive channels.
Conversely, for small mass differences (up to a few hundreds \mevcc), it can be useful to perform a fully 
inclusive search, looking  for a peak in the $\gamma_{ISR}$ energy spectrum, thanks to the good absolute resolution at small photon energies, although this 
kind of searches suffers from a very large background, due to photons from continuum $\epem \to q \bar q$ events.

Some special strategy can be adopted if an energy scan is performed: after reconstructing an exclusive final state, one can apply some selection criteria 
asking for a consistency between the beam and the reconstructed 4-momentum, or conversely avoid such a selection to not exclude ISR or some 
other production mechanisms where part of the event is not reconstructed. In the first case, the event count at each scan step is used to build the 
line shapes of possible resonances. In the second case, where a resonance can be detected at any energy above its mass, one can take the data of all 
scan points together and look for a peak in the invariant mass distribution of the hadronic system. 
Finally, if the center of mass energy can be set at the resonance mass, the production can be strongly enhanced and also the most rare decay channels can be studied.

If a two photon fusion occurs, it is not possible to reconstruct the $\epem$ pair, that is often emitted almost along the beam axis, out of the detector acceptance.
It means that it is not possible to apply any global kinematical constraint. Anyway, these features can be exploited by requiring each particle in the final state to have
a minimum transverse momentum of a few hundreds \mevc, but the whole hadronic system to have a small total transverse momentum, 
typically $P_t \lesssim 50$~\mevc. Then, again, a peak is searched for in the invariant mass distribution of the hadronic system.

At a hadron collider, prompt production mechanisms coexist with the production in $B$ meson decays. Moreover, no beam constraint can be used to cleanly 
identify the $B$ meson. Hence, the standard procedure simply consists in looking for a peak in the invariant mass distribution of some exclusive channel. 
Then, in order to distinguish the prompt and the $B$ components, the impact parameter of the hadronic system is used: the presence of a 
secondary vertex is an indication that a $B$ meson has been produced, flew for a while and then produced 
the hadronic system in its decay. It should be also noticed that triggers play an important role in such searches. In particular, $J/\psi$-oriented trigger 
lines are developed, looking for $\mu^+\mu^-$ pairs (two tracks of opposite charge in the muon system, associated with tracks in the inner tracker).

At a fixed-target experiment like COMPASS, the production mechanisms are the so called \emph{central production}, \emph{diffractive dissociation} 
and \emph{photoproduction}. In the first case, the interaction is described as the emission of two Reggeons, one from the projectile and the other from 
the target nucleus. The fusion of the two Reggeons produces a hadron system where resonances are searched for. In this case, a significative rapidity gap 
is observed between the outgoing  projectile, target and hadron system. Diffractive dissociation and photoproduction are instead described as an interaction 
between the projectile and a Reggeon or photon, emitted by the target. In this case, the event is characterized by a marked forward kinematics, that helps to 
identify this kind of mechanism. Central production is the good place to look for glueballs, while resonances with exotic $J^{PC}$ combinations can be 
produced by diffractive dissociation or photoproduction. Light unflavored mesons\footnote{I follow the PDG conventions and call \emph{light unflavored}
all the mesons with $S = C = B = 0$ and no charm or bottom quark content. The standard quark combinations are $u \bar d$, $(u \bar u - d \bar d)/\sqrt{2}$,
$d \bar u$ and $c_1(u \bar u + d \bar d) + c_2(s \bar s)$.} with a mass up to a few \gevcc and decaying to pions and kaons are typically
searched for in this kind of experiments. In particular, new states appear as resonances in the invariant mass distribution of the hadron system (or part of it). 
Anyway, in this regime, different resonances, or the same resonance through different decay mechanisms, can produce the same final state, and it is important 
to separate the different contributions. One of the most used approaches is the isobar model, where the decay process is described as a succession of two-body 
decays, and the whole process is studied with a Dalitz plot technique, with intermediate resonant states appearing as clear structures in the Dalitz plot. We will 
discuss it in detail in the next sections.

\subsubsection{Measurement of quantum numbers}

The measurement of the quantum numbers is crucial for the determination of the nature of a resonance. In some cases, the
assignment is trivial, being imposed by the production mechanism or the decay channel. For instance, as already mentioned, 
only $1^{--}$ states can be produced via ISR. Similarly, only $C = +$ is allowed by the two photon production, while $J$ and
$P$ follow the selection rules imposed by the so-called Yang's theorem.\cite{yang_theorem} The same arguments
apply to the resonances decaying into photon pairs. In other cases, the production and decay mechanism can 
suggest at least some favored assignment. For instance, only spin-$0$ states have been observed so far in the double charmonium
production along with a $\psi$, so that this assignment is usually the favored one for resonances produced with this mechanism.

When no ultimate indication comes from the production or decay mechanism, an angular analysis has to be performed. It 
can be carried out in different ways, with a different level of detail and difficulty. Let us first
consider the most simple option, which can be applied when there are a few possible assignments and a chain of sequential two-body 
decays (or an isobar model is assumed). In such a case, one angle is chosen, usually the so-called \emph{helicity angle}: for a decay 
chain $X \to Y_1 Y_2$ with $Y_1 \to Z_1 Z_2$ it is defined as the angle between the $Z_1$ (or $Z_2$) momentum in the $Y_1$ rest frame and the 
$Y_1$ momentum in the $X$ rest frame. Then, the distribution of this angle is studied, with theoretical calculations or Monte Carlo 
simulation, for the different quantum numbers hypotheses of $X$. Data are finally compared to the different predictions in 
order to determine the favored one. An example of such a strategy can be found in Ref.~\refcite{Z3930_babar}. 

This is actually a specific application of a more general approach that makes use of the helicity formalism by Jacob and Wick.\cite{jacob_wick}
The basic idea is that, in a reaction like:
\begin{eqnarray}
A \, + \, B \to C \; + & X & \\ 
& X  & \to Y_1 \, Y_2 \, , 
\end{eqnarray}
given the helicities $\lambda_X$, $\lambda_{Y_1}$ 
and $\lambda_{Y_2}$, the decay amplitudes are proportional to the Wigner functions 
$D^{J_X}_{\lambda_X,\lambda_{Y_1}-\lambda_{Y_2}}(\phi,\theta,-\phi)$, being $J_X$ the total angular momentum of $X$
(the formalism can be also extended to sequential two-body decays). So, a parameterization of the angular distributions of $Y_1$ and $Y_2$
can be obtained from a sum of amplitudes $a_{J_i,\lambda_j}$, with definite helicity and total angular momentum hypothesis about $X$:

\begin{equation}
P(\mathbf{\tau}) = \left|\sum_{J_i,\lambda_j} a_{J_i,\lambda_j}\right|^2 = \left|\sum_{J_i,\lambda_j} A_{J_i,\lambda_j} \psi_{J_i,\lambda_j}(\mathbf{\tau})\right|^2 \, ,
\end{equation}
where $\mathbf{\tau}$ indicates the set of angles, and the angular dependences are collected in the functions $\psi_{J_i,\lambda_j}(\mathbf{\tau})$, which include 
the interference among different helicity combinations in the final state.

\begin{figure}[pb]
\centerline{\psfig{file=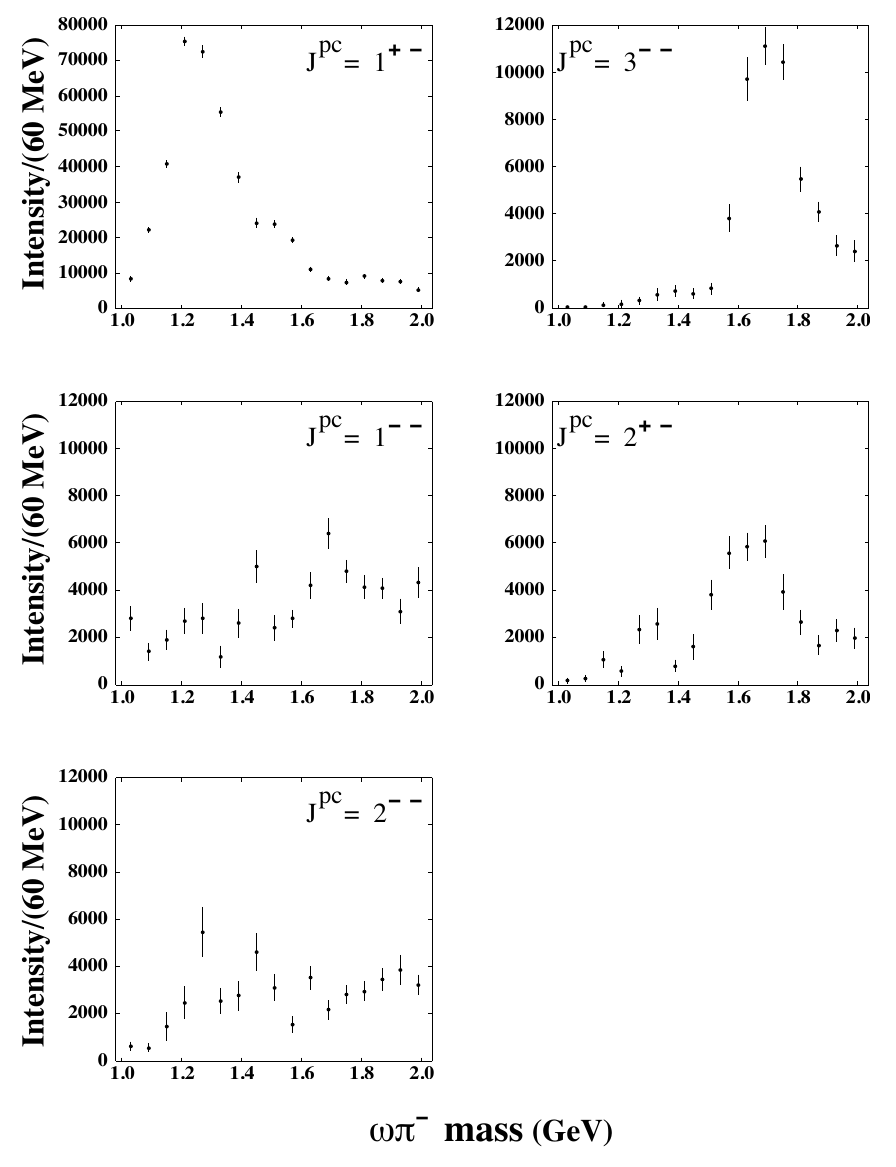,width=10cm}}
\vspace*{8pt}
\caption{Partial wave analysis of $\pi^-p \to \omega \pi^- p$. The $b_1(1235)$ and the $\rho_3(1670)$
are evident in the $1^{+-}$ and $3^{--}$ waves.\label{fig:pwa}}
\end{figure}

By fitting the angular distributions, the partial amplitudes $A_{J_i,\lambda_j}$ can be extracted as a function of the $Y_1 \, Y_2$ invariant mass and, usually, 
the resonance will show up as a peak in some of them, corresponding to the angular momentum $J_i = J_X$. The amplitudes can be also
rearranged so that each combination is also a parity eigenstate. In this case, the peaking amplitude also defines the $X$ parity. An example
of the results of such an analysis in shown in Fig.~\ref{fig:pwa}, that is taken from Ref.~\refcite{pwa}.

If a three body decay $X \to Z_1 \, Z_2 \, Y_2$ is studied, but an isobar model is used, with $X \to Y_1 \, Y_2$, $Y_1 \to Z_1 \, Z_2$, it is possible to use different 
waves for the different choices of $Y_1$ and the different angular momenta of $Y_1$ versus $Y_2$, by fitting also the $Z_1 \, Z_2$ invariant mass distribution
(e.g. in the $X \to \pi \pi \pi$ decay, one can associate an amplitude to the $P$-wave $X \to \rho \pi$, one to the $S$-wave $X \to f_0 \pi$, etc.). 
In this case, each amplitude has also definite isospin ($I$) and $C$-parity,\footnote{When the hadron system is not neutral, is it still possible to define
the $C$-parity by assign to charged mesons the $C$-value of the neutral member of their multiplet.} and the resonance will show up only in the 
amplitudes having $C = C_X$ and $I = I_X$. Due to the description of the decay amplitude as a sum of different $J^{PC}$ waves, the technique is usually 
referred to as a \emph{partial wave analysis}.

In some cases, additional information can be extracted from the invariant mass distribution of the $Y_1 \to Z_1 \, Z_2$ system, in particular if $X$ 
is near the threshold for the production of this pair, because in this case threshold effects arise, that are different depending 
on the orbital angular momentum of $X$.

\subsubsection{Background description}
\label{sec:background}

In most cases, resonances clearly appear as an invariant mass peak on top of an almost flat background. In this case, an empirical description of 
the background distribution
is enough to guarantee a reliable result. Anyway, in some cases one need to be more careful. A good example is the search for the charged state $Z(4430)^\pm$ at
Belle\cite{Z4430_belle} and BaBar\cite{Z4430_babar}. I discuss this benchmark case in some detail. 

\begin{figure}[pb]
\centerline{\psfig{file=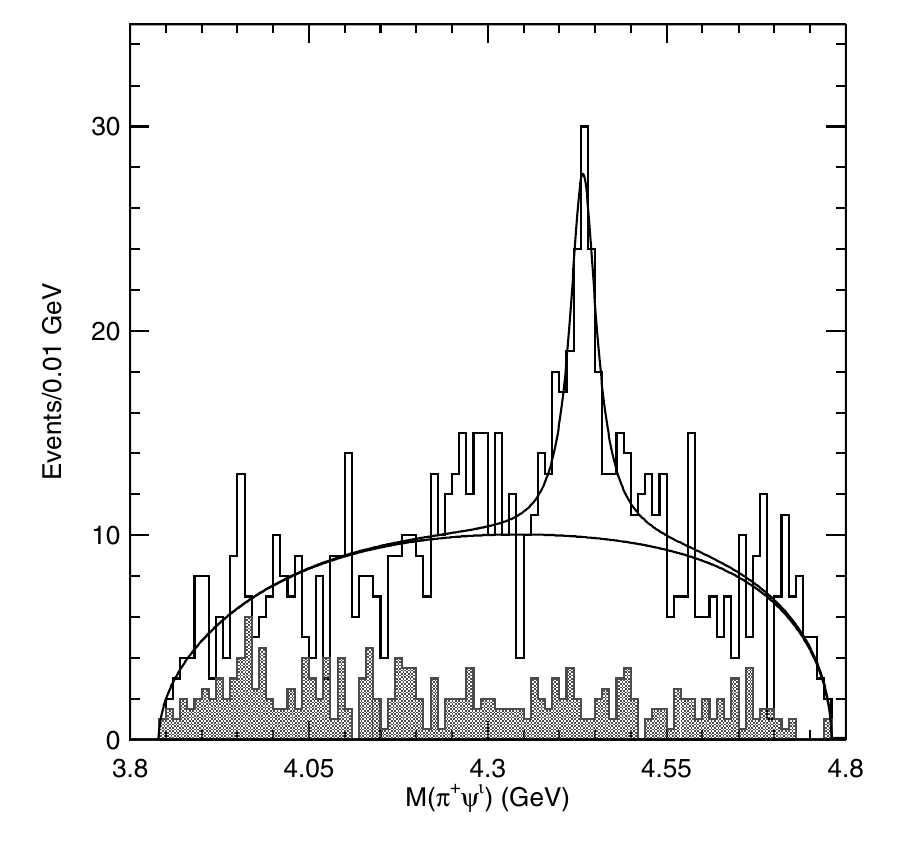,width=7cm}}
\vspace*{8pt}
\caption{The $\psi(2S)\pi^\pm$ invariant mass distributions in $B \to K \psi \pi^\pm$ as seen by Belle. \label{fig:4430_belle}}
\end{figure}

\begin{figure}[pb]
\centerline{\psfig{file=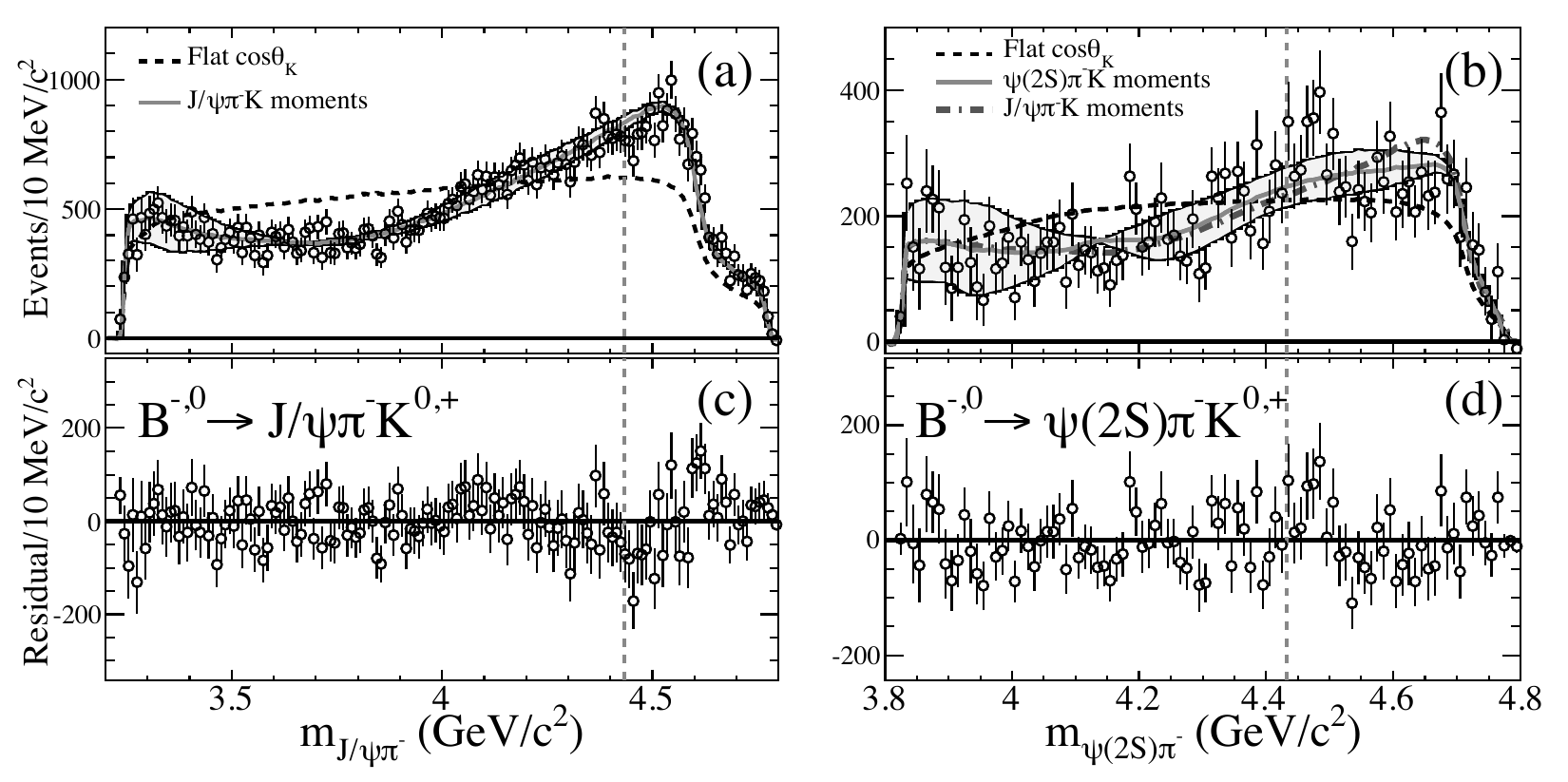,width=12cm}}
\vspace*{8pt}
\caption{The $J/\psi \pi^{\pm}$ and $\psi(2S)\pi^\pm$ invariant mass distributions in $B \to K \psi \pi^\pm$ as seen by BaBar. \label{fig:4430_babar}}
\end{figure}

A charged state at 4.43~\gevcc was firstly claimed by Belle looking at the $\psi(2S)\pi^\pm$ invariant mass distribution in the $B \to K \psi(2S)\pi^\pm$ decay,
with an empirical description of the background (see Fig.~\ref{fig:4430_belle}). Anyway, in this case, the background distribution turns out to be quite complicated, 
as shown in Fig.~\ref{fig:4430_babar}. This distribution has been explained by the BaBar collaboration as a consequence of the resonant structures 
in the $K \pi^{\pm}$ system. BaBar showed that resonant $P$-wave contributions in the $K \pi^{\pm}$ system (mainly the $K^*(892)$) interfering 
with non resonant $S$-wave contributions generate significant asymmetries in the polar angular distribution of kaons with respect to the $K \pi^{\pm}$ flight 
direction. As a consequence, also the $\psi(2S)\pi^\pm$ invariant mass distribution is strongly deformed. To take it into account, 
BaBar performed a Dalitz plot analysis: at first, they give a phenomenological description of the $K \pi^{\pm}$ spectrum by a sum of 
$S$-, $P$- and $D$-wave contribution, where resonant contributions are described by Breit-Wigner amplitudes and continuum contributions 
are modeled on the experimental results for $K-\pi$ scattering; then, the angular distribution is described with a partial-wave formalism, with
a dependence on the invariant mass of the $K \pi^{\pm}$ system. As a result, BaBar did not confirm the Belle discovery, but found an hint of a 
peak in the $\psi(2S)\pi^\pm$ invariant mass distribution with a mass significantly lower that the one reported by Belle. By repeating its analysis 
with a similar technique\cite{Z4430_belle_new}, the Belle collaboration confirmed its previous result, but the uncertainty on the mass 
of the resonance increased significantly, making it consistent with the position of the peak glimpsed by BaBar.

This discussion illustrates that, when the background distribution shows some clear structure and there is a hint of a possible influence of interfering 
channels, it can be necessary to attain a phenomenological description of the background, instead of providing simple empirical parameterizations.

\section{Review of observed exotica}
\label{sec:states}

In this section I will review most of the exotic states discovered in the last few years. Again,
I will concentrate on some experimental aspects, the theoretical issues being already discussed 
in several review papers I already referenced to. Also, I do not pretend to give an exhaustive
overview, but instead concentrate on a few states and aspects that clarify how the experimental
studies guide the theoretical interpretation of the exotic states.

As I stressed in Sec.~\ref{sec:theo}, it is of paramount importance to identify possible multiplets in
the new hadron zoology. Hence, whenever possible, I will classify the new resonances in families, 
based on their mass and quantum numbers, although there is no definitive indication yet
to identify these families as multiplets.

\subsection{Light exotic states}

In this section I discuss a few light unflavored states that escape a standard description as $q \bar{q}'$ pairs The first of the
list are the $a_0(980)$ and the $f_0(980)$, that are scalar mesons ($J^{PC} = 0^{++}$) with isospin 1 and 0, respectively. 
They do not fit in the standard picture for a series of reasons: they are 100~\mevcc below the mass predictions for the $1\,^3P_0$ states, 
their coupling to $\pi\pi$ and $\gamma\gamma$ are too small and, conversely, the coupling to $K \bar K$ are too large.
Several interpretations have been proposed for these states, including tetraquarks and $K \bar K$ molecules. What is
sure is that the $K \bar K$ component must be large, but it does not imply a molecular composition: it has been shown\cite{980_KK} that
such a component can be obtained when the quarks recombine at large distances before the decay. For a detailed review of the 
possible interpretations and a list of references, see Ref.~\refcite{pdg}.

The possibility of having light tetraquark states is strengthen also by some observations made at
the LEP collider. In particular, the measurement of the $\rho \rho$ production cross section in two photon processes
could suggest the presence of an exotic isospin 2 resonance around 1.5~\gevcc, which could be
identified as a tetraquark.\cite{pire}

\begin{figure}[pb]
\centerline{\psfig{file=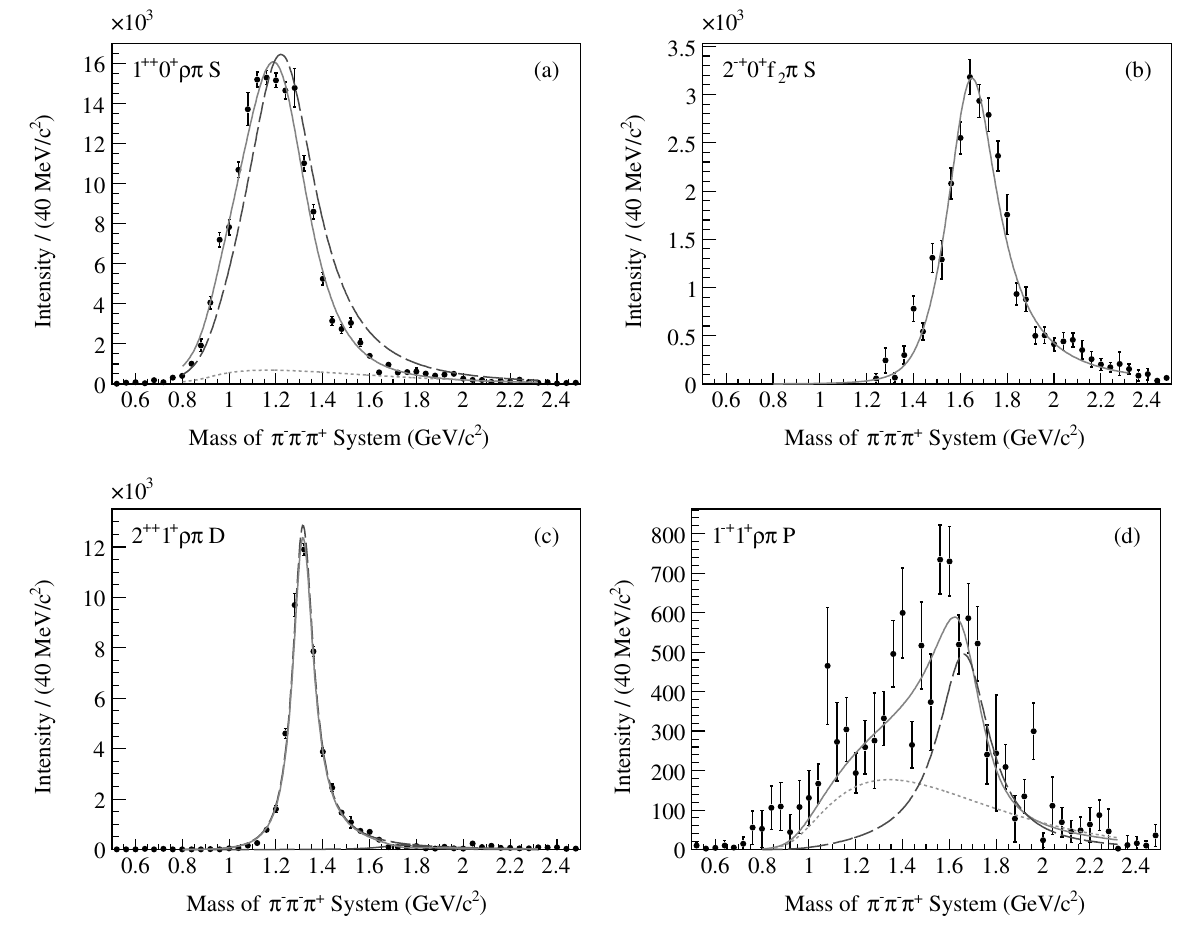,width=12cm}}
\vspace*{8pt}
\caption{Main partial waves for the $\pi^- p \to \eta' (\pi^+\pi^-) \pi^- p$ process at COMPASS. The waves with standard quantum number combinations
$1^{++}$, $2^{-+}$ and $2^{++}$ show the contributions from the well-known $a_1(1260)$, $\pi_2(1670)$, and $a_2(1320)$, respectively, with some 
minor resonant and non-resonant components. The $\pi_1(1600)$ shows up as a peak in the $1^{-+}$ $\rho\pi$ $P$-wave
over a non-resonant background. \label{fig:pi1600_pwa}}
\end{figure}

In the recent past, new information and discoveries came from COMPASS, BES-III and CLEO-$c$. In particular, from a data taking 
with a 190~\gev pion beam on a liquid $\rm{H}_2$ target, the COMPASS collaboration confirmed the so-called $\pi_1(1600)$,\cite{pi1600_compass}
observing a strong signal at 1.66~\gevcc in the three pion mass distribution of the $\pi^- p \to \eta' (\pi^+\pi^-) \pi^- p$ diffractive dissociation reaction. 
As shown in Fig.~\ref{fig:pi1600_pwa}, a partial wave analysis indicates that the signal is produced by a $1^{-+}$ state, an exotic 
quantum number combination that makes it a very good hybrid candidate. 
A further confirmation of this state came from a preliminary result by CLEO-$c$, that observed it with a 4$\sigma$ significance in the 
$\eta' \pi$ invariant mass spectrum of the $\chi_{c1} \to \eta' \pi^+ \pi^-$ decay.\cite{pi1600_cleo} This is the first time that this structure 
is observed in charmonium decays, which tends to confirm its resonant nature in contrast with a possible interpretation as a threshold effect.

\subsection{Exotic charmonium-like states}

\subsubsection{The X(3872)}

\begin{figure}[pb]
\centerline{\psfig{file=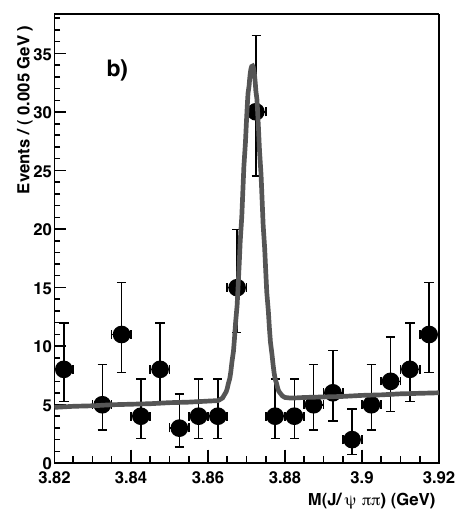,width=6cm}}
\vspace*{8pt}
\caption{The $X(3872)$ signal in the $J/\psi \pi^+ \pi^-$ invariant mass distribution by Belle. \label{fig:X3872_belle}}
\end{figure}

As already mentioned, the $X(3872)$ resonance is the forerunner of the new charmonium spectroscopy. It was 
discovered by Belle\cite{X3872_belle} (see Fig.~\ref{fig:X3872_belle}) in the $B \to K X$ decay, with $X \to J/\psi \pi^+ \pi^-$, soon confirmed
by BaBar\cite{X3872_babar} and CDF\cite{X3872_cdf}, and now by LHC.\cite{X3872_lhcb,X3872_cms} It was soon observed also 
in the $X \to D \bar {D}^*$ channel, with the interesting feature that there is a tension (3.5$\sigma$) between the 
masses measured in the two decay channels. This suggested the possibility of having two different and near states, 
as predicted for a tetraquark. The angular analysis by CDF also allowed to exclude quantum numbers other than $1^{++}$
and $2^{-+}$, where $C = +$ is also confirmed by the observation of the $X \to J/\psi \gamma$ decay.

This state immediately appeared to be a good exotic hadron candidate: at first, although it is above the
open-charm threshold, its total width is much smaller than what is expected for a regular charmonium:
the measured width\cite{Gamma3872_babar} is $(3.0^{+1.9}_{-1.4} \pm 0.9)$~\mev, to be compared with a typical width of $\sim$30~\mev for 
regular charmonia of similar mass. Moreover, at the time of its discovery, the favored quantum numbers were $J^{PC} = 1^{++}$, like the $\chi_{c1}(1P)$, but 
the mass is far from the predictions for any radial excitation of this resonance. Another strong indication of the exotic nature
of the $X(3872)$ comes from the evidence of a $X \to J/\psi \rho$ dominance in the $X \to J/\phi \pi^+ \pi^-$ decay\cite{X3872_cdf}, that would
imply isospin violation if the $X(3872)$ be a regular charmonium. 

\begin{figure}[pb]
\centerline{\psfig{file=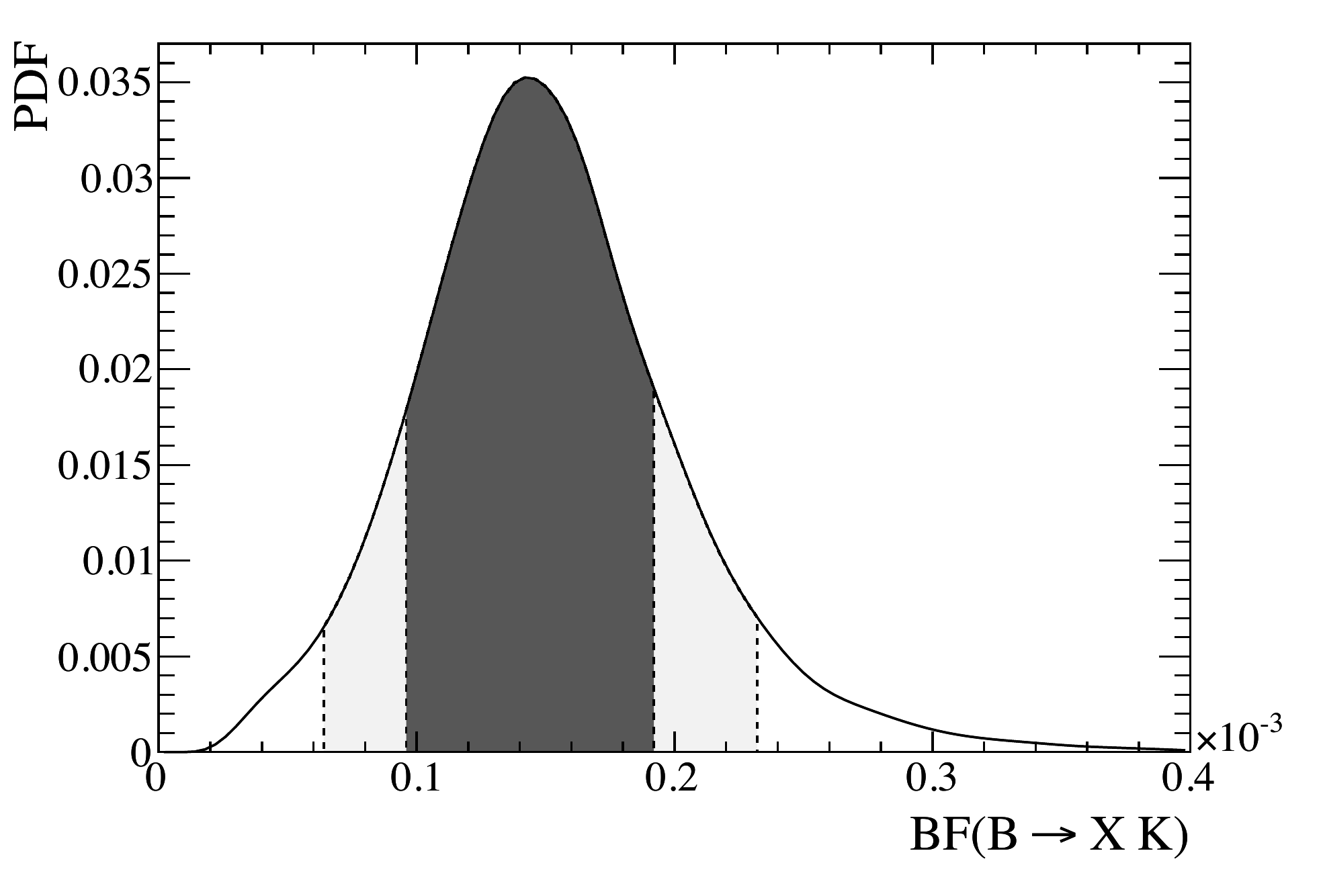,width=6cm} \psfig{file=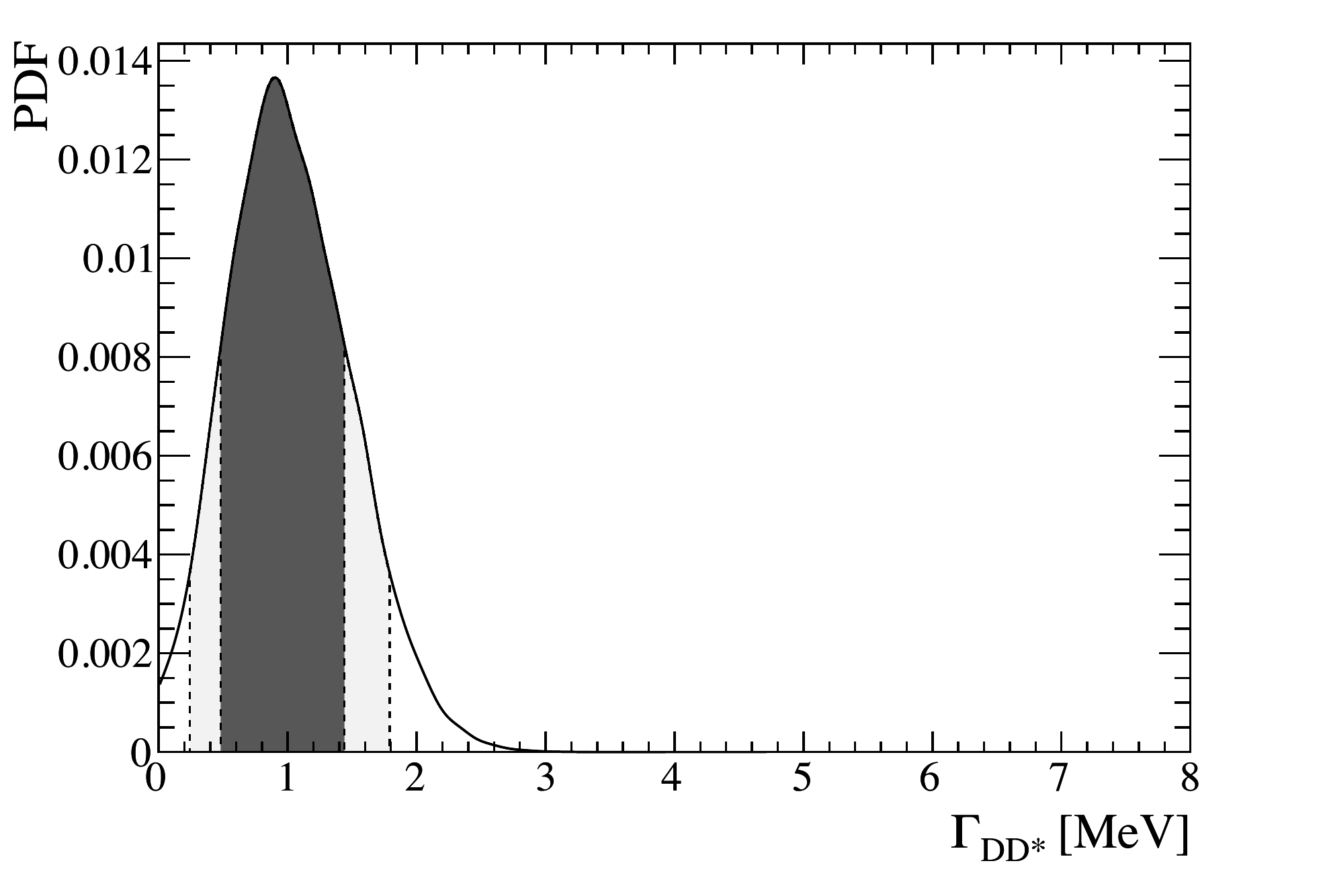,width=6cm}}
\vspace*{8pt}
\caption{Total BR of $B \to K X$ and partial width $\Gamma(X \to D \bar{D}^*)$ from a bayesian combination of available measurements  
of the observed $X(3872)$ decays. \label{fig:X_bayes}}
\end{figure}

The $X(3872)$ is also the most studied of all new resonances. Several decay channels have been searched for and studied, and a
combination of the experimental results has been reported in Ref.~\refcite{reviewNC}. Among the outcomes of this study, two deserve
a discussion here (see Fig.~\ref{fig:X_bayes}). At first, the total BR of $B \to X K$ is found to be $\sim 10^{-4}$, which would be unusually low for
a regular charmonium, that typically gives values $\sim 10^{-3}$: this is another important argument in favor of an exotic interpretation.
Second, $\Gamma(X \to D \bar {D}^*)$ is found to be $\sim 1$~\mev, and I explained in Sec.~\ref{sec:discriminants} that a width of $\sim 70$~\kev
is the most natural value for a $D \bar {D}^*$ molecule: this result disfavors a molecular interpretation of this resonance.

Several works have been also devoted to the possible presence of two near states, as suggested by the comparison of the invariant mass spectra of
$X \to J/\psi \pi^+ \pi^-$ and $X \to D \bar {D}^*$. It can be further tested looking for a mass difference between the states observed in $B^0 \to K^0 X$ and 
$B^\pm \to K^\pm X$, when $X$ decays into the same channel. Alternatively, multiple structures can be searched for in the invariant mass spectrum. 
The second option has been investigated by CDF,\cite{X3872_CDF_mass} which excludes at 90\% confidence level (C.L.) a mass difference larger than 3.2~\mevcc. 
The first one has been studied by BaBar and Belle,\cite{X3872_babar_mass,X3872_belle_mass} which also excluded a difference at $\sim 1$~\mevcc level.

A couple of other results recently added some important indication about the nature of the $X(3872)$. BaBar realized an angular
analysis of $X \to J/\psi \omega$ ($\omega \to \pi^+ \pi^- \pi^0$), and found that $J^{PC} = 2^{-+}$ is actually favored with respect to
$1^{++}$: this result makes possible an interpretation as a regular charmonium: the $\eta_{c2}(1D)$. Both BaBar\cite{X3872_psigamma_babar} and 
Belle\cite{X3872_psigamma_belle} searched for the the $X \to J/\psi \gamma$ and $X \to \psi(2S) \gamma$ decays, that are predicted to dominate 
for a molecule. Although there is some tension between the two results, the possibility of a $\psi \gamma$ dominance seems to be excluded. Finally, a very recent
report by Belle\cite{X3872_belle_updates} clarified a few questions. At first, it confirmed that the mass difference between the states produced in $B^0 \to K^0 X$ and 
$B^\pm \to K^\pm X$ is excluded at the level of 1~\mevcc, in contrast with the hypothesis of two different tetraquark states forming a multiplet. Second,
there is no evidence for a charged state at a similar mass, decaying to $J/\psi \pi^+ \pi^0$: it indicates that the $X(3872)$ is most probably an
isospin singlet. Unfortunately, an attempt to discriminate between $J^{PC} = 1^{++}$ and $2^{-+}$ with an angular analysis of $X \to J/\psi \pi^+ \pi^-$
didn't brought to any conclusion, due to the lack of statistics.

In conclusion, there are several arguments against a molecular interpretation, while a tetraquark interpretation is favored by some observations, but it is
not confirmed by other measurements explicitly performed for this purpose. Moreover, the $2^{-+}$ assignment by BaBar reopened the possibility of a 
standard interpretation.

\subsubsection{The 3940 family}

Probably the most interesting group of new resonances of similar masses is the so-called 3940 family. It is composed by four
states lying around 3.94~\gevcc, produced with different mechanisms and observed in different final states. They were
all discovered by the Belle collaboration.

The first resonance to be discovered among them is the $Y(3940)$, produced in the $B \to K Y$ decay and observed in 2005 in
the $J/\psi \omega$ final state.\cite{Y3940_belle} One year later, a resonance was found in the two photon fusion process, looking at
the $D \bar D$ final state,\cite{Z3930_belle} and it was called $Z(3930)$. Later, a new state was obtained with a double charmonium 
production, $\epem \to J/\psi X$, and observed in the $D \bar {D}^*$ final state:\cite{X3940_belle} the X(3940).
Finally, in 2010 BaBar\cite{Y3915_babar} observed also the so-called $Y(3915)$, a two photon fusion resonance decaying into $J/\psi \omega$.

For both the $Y(3940)$ and the $Y(3915)$ the only quantum number determined so far is $C = +$, following from the mechanism production
and the decay channel, respectively. Moreover, the still large uncertainty ($\sim$ 20\mev) in the determination of their mass makes possible to identify them 
as a single state. It would be the first of the new resonances observed with different production mechanism. 
One could also speculate that the states produced in double charmonium production (the $X(3940)$) and $B$ decays (the $Y(3940)$) be the same resonance.
Anyway, this hypothesis is disfavored by the observation that  $BR(X \to J/\psi \omega)/BR(X \to D \bar{D}^{*}) > 0.58$, while
$BR(Y \to J/\psi \omega)/BR(Y \to D \bar{D}^{*}) > 0.71$, both at 90\% confidence level.

The preferred assignment for the $X(3940)$ is $J = 0$, given that all states observed up to now in the double charmonium production are spin 0.
The $X(3940)$ also has positive charge conjugation, as imposed by the two photon fusion production. Hence, the preferred $J^{PC}$ combinations are 
$0^{++}$ and $0^{-+}$. 

Finally, BaBar studied the angular distribution of $Z(3930) \to D \bar D$ and found that $J = 2$ is the preferred 
assignment,\cite{Z3930_babar} while only $C = +$ is allowed in the double photon production. It makes possible an identification of this 
state with the regular $\chi_{c2}(2P)$.

\subsubsection{The 4140 family}

Another interesting group is the 4140 family, composed by two resonances: the $X(4160)$, produced through a double charmonium production and decaying to
$D^* \bar {D}^*$, and the $Y(4140)$, produced in $B$ decays and observed in $J/\psi \phi$. They were discovered by Belle and CDF,\cite{X4160_belle,Y4140_cdf}
respectively.

The possible $J^{PC}$ assignments for the $X(4160)$ are $0^{\pm+}$ and $2^{\pm+}$, and the first hypothesis could suggest an identification with the $\eta_c(3S)$.
For the $Y(4140)$, along with the standard $0^{++}$ and $2^{++}$ combinations, the exotic $1^{-+}$ assignment is possible, as expected for some hybrids.
In order to test this hypothesis, the $Y(4140)$ has been searched for in the double photon production, that is expected to be enhanced for hybrid hadrons.
Unfortunately, no evidence for such resonance has been found there, but a new state was discovered, called $X(4350)$. Moreover, analyzing 
the $B \to K J/\psi \phi$ channel with more statistics, CDF also found some indication for a new state at 4.274~\gevcc.

\subsubsection{The $1^{--}$ family}

At \epem colliders, the $1^{--}$ states are among the most easiest to be found, since they can be extensively produced through the ISR mechanism. Four new
states were discovered at the \Bfactories:\cite{Y4008,Y4260,Y4350,Y4660} the $Y(4008)$, the $Y(4260)$, the $Y(4350)$ and the $Y(4660)$. The 
first two decay to $J/\psi \pi^+ \pi^-$, while the second two decay to $\psi(2S) \pi^+ \pi^-$. The $Y(4260)$ has been also confirmed by 
CLEO,\cite{Y4260_cleo,Y4260_cleo_dir} that could copiously produce it by running at $\sqrt{s} \sim 4.26~\gev$. This allowed to 
observe also the $Y \to J/\psi \pi^0 \pi^0$ and the $Y \to J/\psi K^+ K^-$ modes.

\begin{figure}[pb]
\centerline{\psfig{file=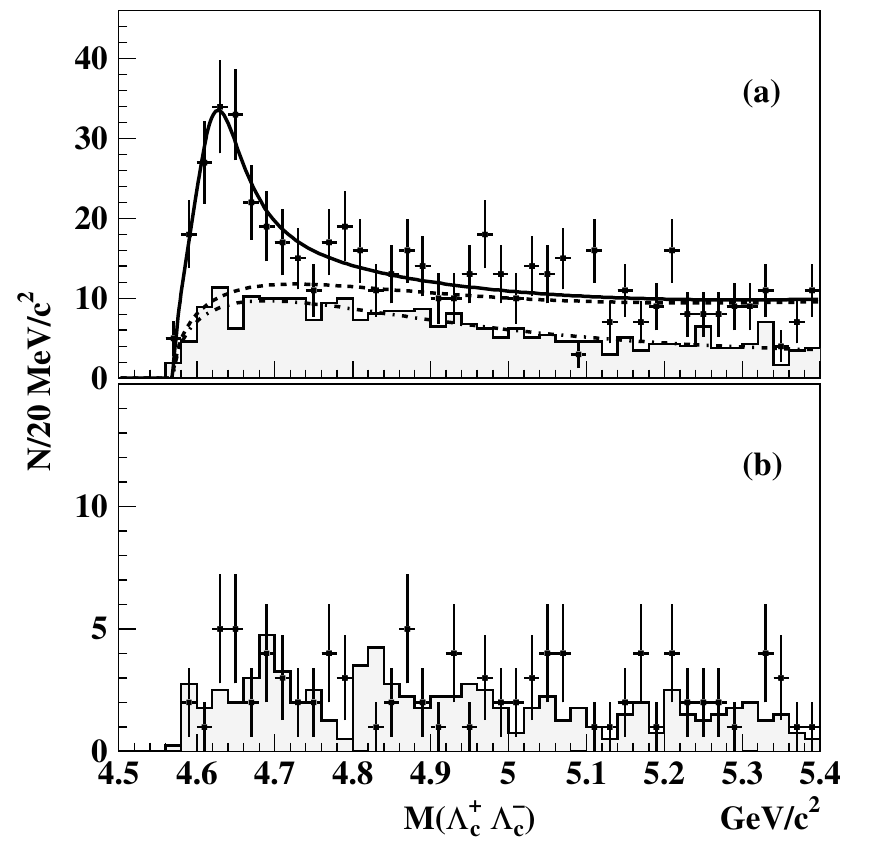,width=7cm}}
\vspace*{8pt}
\caption{The $Y(4660) \to \Lambda_c \Lambda_c$ signal as observed by the Belle collaboration (a) and a validation of the background model in a 
control sample (b). \label{fig:Y4660_LL}}
\end{figure}

Searches have been performed to look for these states decaying into \DDbar or the baryonic mode 
$\Lambda_c \Lambda_c$: the former is the preferred channel for regular charmonia, while the latter should be enhanced for tetraquarks. No evidence
has been found so far for the \DDbar final state, setting upper limits at the level of unit for 
$\Gamma(Y \to D \bar D)/\Gamma(Y \to J/\psi \pi^+ \pi^-)$,\cite{Y4260_DD} while a large $\Lambda_c \Lambda_c$ signal was 
found for the $Y(4660)$, as shown in Fig.~\ref{fig:Y4660_LL}, with $BR(Y \to \Lambda_c \Lambda_c)/BR(Y \to  J/\psi \pi^+ \pi^-) = 25 \pm 7$.\cite{Y4660_LL} 
At present, it is the best of all tetraquark candidates.

An analysis has been also performed by Belle to compare the $Y(4260) \to  J/\psi \pi^+ \pi^-$ and the 
$Y(4260) \to  J/\psi \pi^0 \pi^0$ rates. Isospin symmetry predicts the second one to be half the first one for standard states, while exotic states could 
violate this rule. At present, no evidence of such a violation has been found, although the uncertainty on the ratio is quite large (60\% level) with the present
statistics.

\subsubsection{Charged states}

As already mentioned, charged states are very interesting because they cannot be obtained as standard charmonia, and in particular a lot of such
resonances is expected under the tetraquark hypothesis.

Searches for charged states have been carried on looking at several final states, including $J/\psi \pi^+ \pi^0$, and Belle\cite{Z4430_belle,Z12_belle} 
observed at the end three states, the $Z(4430)^+$, the $Z_1(4050)^+$ and the $Z_2(4250)^+$, the first decaying into $\psi(2S) \pi^+$ and the others into 
$\chi_{c1} \pi^+$. These states have not yet been confirmed by other experiments, and I already discussed the problematic treatment
of the background in Sec.~\ref{sec:background}.

\subsection{Exotic bottomonium-like states}

\begin{figure}[pb]
\centerline{\psfig{file=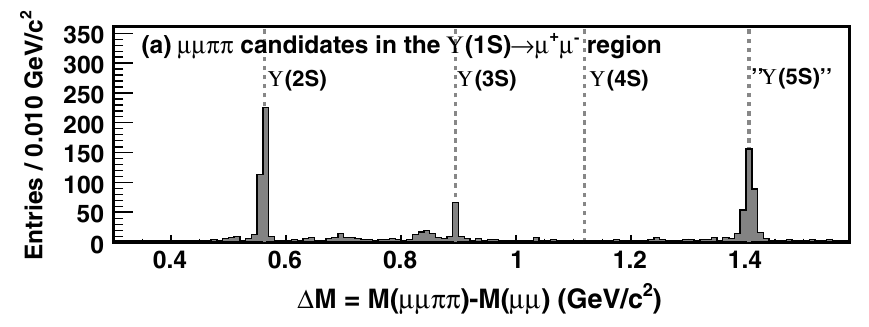,width=9cm}}
\vspace*{8pt}
\caption{$\Upsilon(1S) \pi^+ \pi^-$ invariant mass spectrum from the Belle pioneering run at the $\Upsilon(5S)$ resonance. \label{fig:Ypipi}}
\end{figure}

The observation of several exotic charmonium-like states motivated the search for their bottom companions. Given that the exotic charmonia appear
above the open charm threshold, the discovery of similar bottomonium-like states at a \Bfactory requires to run it at a higher energy with respect to 
the normal operations, performed at the $\Upsilon(4S)$ (which is just above the \BBbar threshold). The first attempts included a pioneering run by 
Belle at the $\Upsilon(5S)$ resonance\cite{belle_5S} and an energy scan between the \BBbar threshold and 11.2~\gev by BaBar,\cite{babar_scan} looking 
for a direct signal of $\epem \to Y_b$. No evidence for new resonances was found, but an intriguing, large 
$\Upsilon(5S) \to \Upsilon(nS) \pi^+ \pi^-$ ($n$ = 1, 2) signal was observed by Belle (See Fig.~\ref{fig:Ypipi}). It was unexpected if compared 
to the similar decays of the $\Upsilon(4S)$ and looked interesting, being the bottom partner of $Y \to \psi(nS) \pi^+ \pi^-$. 

\begin{figure}[pb]
\centerline{\psfig{file=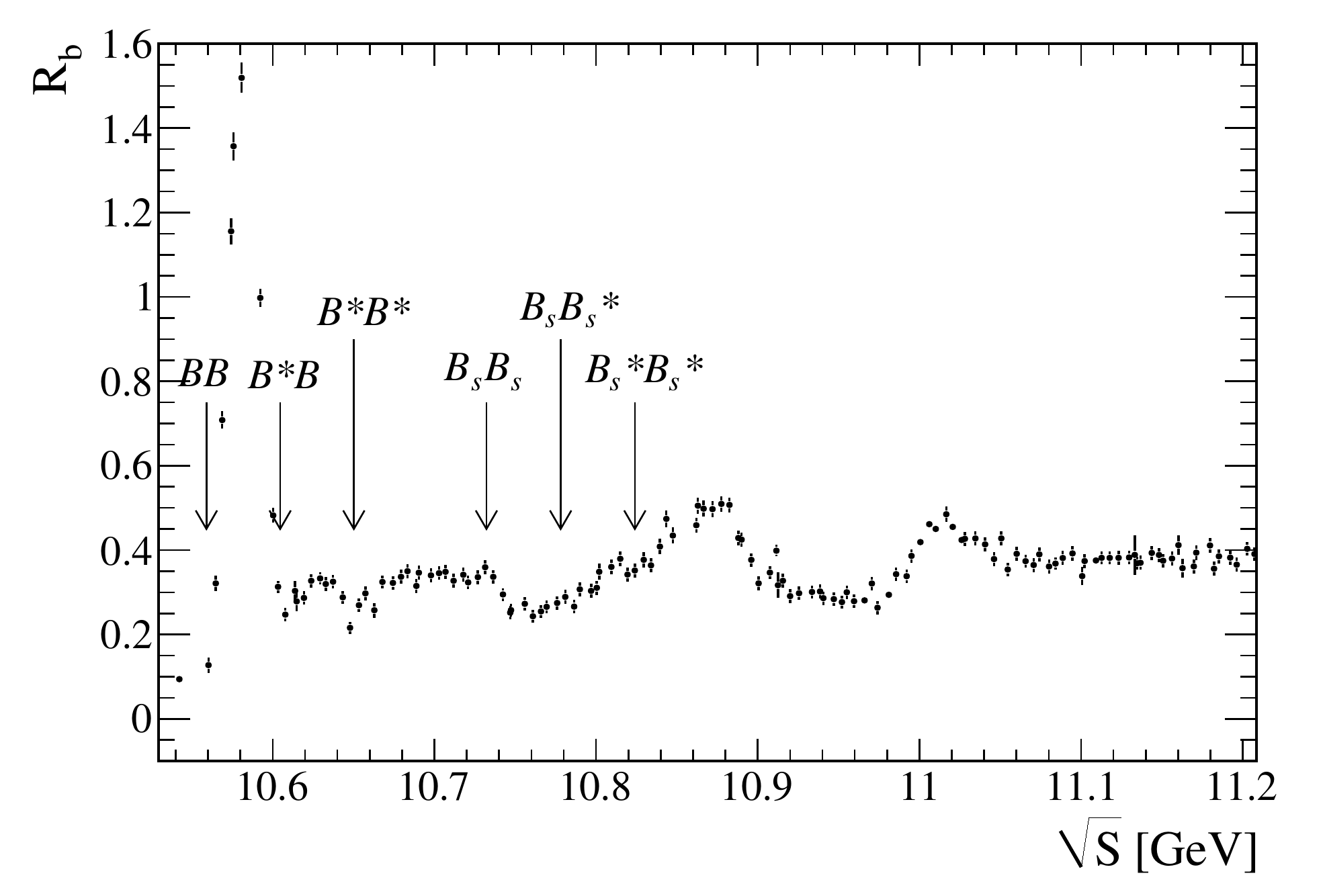,width=6cm} \psfig{file=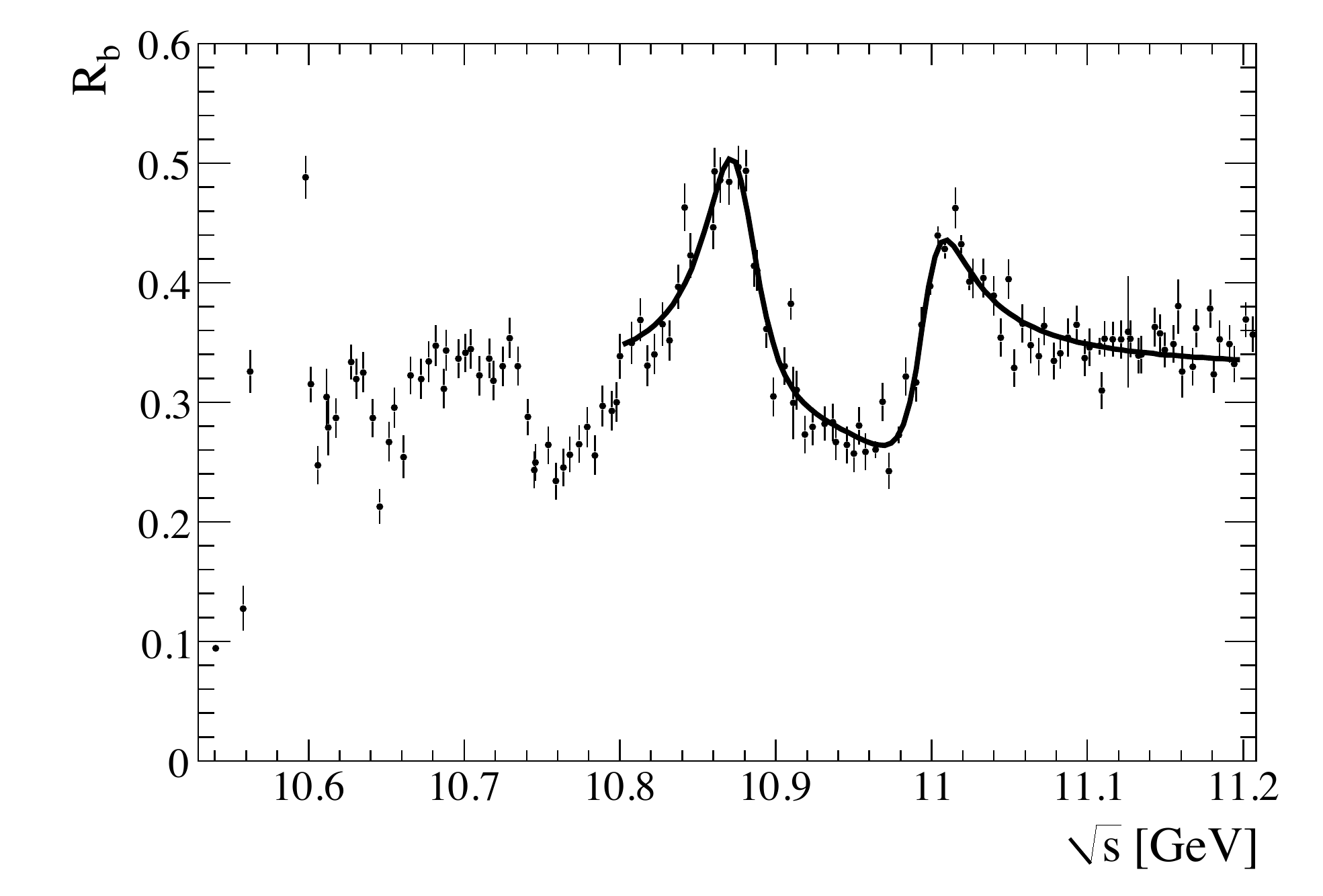,width=6cm}}
\vspace*{8pt}
\caption{Measurement of the $\epem \to b \bar b (\gamma)$ cross section by BaBar between 10.58 and 11.2~\gev, with the position of the different
$B \bar B$ thresholds (left). The region around the $\Upsilon(5S)$ and the candidate $\Upsilon(6S)$ resonances is fitted with the model described 
in the text (left). \label{fig:xsec}}
\end{figure}

Triggered by this first observation, Belle performed a scan around the $\Upsilon(5S)$, for a total of $\sim 7$~\invfb of integrated luminosity, 
and found a different peak position in the inclusive $\epem \to hadrons$ rate with respect to the exclusive $\Upsilon(nS) \pi^+ \pi^-$ channels,\cite{belle_5S_scan} 
arguing that a new exotic resonance could lie just near the standard $\Upsilon(5S)$. This result was obtained by fitting the cross section of the inclusive 
and exclusive modes, as a function of $\sqrt{s}$, with a simple Breit-Wigner shape interfering with a flat continuum. 
A mass of $(10.8884^{+0.0027}_{-0.0026} \pm 0.0012)~\gevcc$ and a width of $(30.7^{+8.3}_{-7.0} \pm 3.1)~\mev$ were measured in the exclusive modes, 
3$\sigma$ and 5$\sigma$ away from the $\Upsilon(5S)$ PDG values,\cite{pdg} which
are instead confirmed by the Belle data with the fit to the inclusive spectrum.
Anyway, the interpretation of this result is controversial. Actually, it is well known\cite{eichten,tornqvist} that final state interactions can modify the 
exclusive and inclusive shapes in different way, producing this kind of apparent discrepancies. The difficulty in the interpretation of this result
is confirmed by the analysis of the BaBar energy scan. This collaboration collected a total of $\sim 3.3$~\invfb of integrated luminosity in the region between
10.54 and 11.2~\gev, in finer steps (5~\mev) with respect to Belle. Then, the inclusive $\epem \to b \bar b (\gamma)$ cross section was measured, 
and fitted in the region between 10.58 and 11.2~\gev, where the $\Upsilon(5S)$ and the candidate $\Upsilon(6S)$ resonances lie. The fit,
whose result is shown in Fig.~\ref{fig:xsec}, was performed with two Breit-Wigner resonances, an interfering and a not-interfering flat continuum, 
$\sigma = |A_{ni}| + |A_{i} + A_{5S}BW(M_{5S},\Gamma_{5S}) + A_{6S}BW(M_{6S},\Gamma_{6S})|$. The results are in disagreement with the PDG values and 
in better agreement with the Belle parameters for the exclusive modes. It was also stressed that large systematic uncertainties are related to the choice
of the specific parameterization of the continuum contributions. Moreover, several structures are evident below the $\Upsilon(5S)$ mass. These structures,
predicted in Ref.~\refcite{tornqvist}, are not associated to any new resonance, but are produced by final state interactions, which confirms the
drawback of any naive interpretation for the inclusive cross section. A new analysis of Belle data with a similar fit approach confirmed these results, 
reducing the discrepancy between the inclusive and exclusive shapes to 2.2$\sigma$ (9 MeV) and $1.4\sigma$ in mass and width respectively.
In conclusion, in absence of a solid model for the $\epem \to \Upsilon(nS) \pi^+ \pi^-$ process and an accurate measurement of the exclusive 
$\epem \to B^{(*)} \bar {B}^{(*)}$ modes,\footnote{Such a precise measurement could come from the 
forthcoming Super \Bfactories. According to the current estimates,\cite{superB_white} one week of data taking at a luminosity of $10^{36}$~\invcminvs would 
make possible this measurement with a 10\% error.} to be used to tune the $\epem \to hadrons$ predictions, it is difficult to draw any conclusion. 
Moreover, the hypothesis of a single state, resulting from the mixing between the standard $\Upsilon(5S)$ and an exotic $Y_b$, is also fascinating.

Recently, Belle decided to perform a long data taking at the $\Upsilon(5S)$, collecting up to 121~\invfb of integrated luminosity. This data set allowed to 
search for exotic states with a much higher sensitivity. In particular, while studying the $\Upsilon(5S) \to h_b(nP) \pi^+ \pi^-$ decay, two structures
were found\cite{Zb_belle} in the $h_{b} \pi^+$ invariant mass distribution, at 10.61 and 10.65~\gevcc. Several interpretations
have been proposed for these two structures, called $Z_{b1}$ and $Z_{b2}$, including $B^{(*)} \bar {B}^{(+)}$ molecules\cite{Zb_molecule} 
(the two structures are near the $B \bar {B}^*$ and the $B^* \bar {B}^*$ thresholds), tetraquarks\cite{Zb_ali} and threshold effects.\cite{Zb_cusp}

\section{Conclusions}

In this paper I described some of the experimental techniques behind the discovery of new resonances, that do not fit in the standard picture of mesons and baryons.
I also gave a review of the exotic candidates observed so far, discussing in particular the experimental observables that can help to establish their nature.

Even if a lot of progresses have been made in the last few years, the situation is far from being completely clarified. In most cases, new data are needed, and they are
expected to come from the LHC and the future flavor factories. Anyway, I also have to point out that the potentiality of present data probably has not yet been 
completely exploited. On the other hand, the analysis effort of the \Bfactory experiments is still on going and new results are expected to come in the near future. Finally, in
some cases, there are theoretical aspects that need to be further investigated, for instance in the case of the $\Upsilon(5S \to \Upsilon(nS) \pi \pi$ interpretation and its
connection with the $\epem \to hadrons$ cross section. Anyway, also in these cases an important help can derive from new or more precise measurements.

In conclusion, hadron spectroscopy is still an intriguing field, that will continue to provide, in the near future, important information to better understand QCD and its
effective treatments, with a broad impact on many other fields. 

\section*{Acknowledgments}
The author would like to thank Professor Riccardo Faccini for its continuous support and for having reviewed the draft of the manuscript.

\appendix

\section{Naming conventions for standard quarkonia}
\label{app:naming}

Quarkonium is a bound state of a quark and its anti-quark, $q \bar q$. Due to the
isospin symmetry, the lightest quark $u$ and $d$ form mixed states like
$\pi^0 = (u \bar u - d \bar d)/\sqrt{2}$. Conversely, the heavy quarks $c$ and $b$ 
form pure charmonium $c \bar c$ and bottomonium $b \bar b$, respectively.
An intermediate situation is found for the strange quark, that can form pure $s \bar s$
states or mix with the $u \bar u + d \bar d$ states. Finally, the short lifetime of the top 
quark prevent the formation of the toponium.

Concentrating on charmonium and bottomonium, the easiest to observe states are the $J^{PC} = 1^{--}$
resonances, being $J$ the total angular momentum, which can be produced through an $s$-channel \epem interaction. 
They are named $\psi$ for charmonium and $\Upsilon$ for bottomonium (with the only exception of the lowest $1^{--}$ 
charmonium state, called $J/\psi$ for well known historical reasons). A principal quantum number $n$ for radial excitation
and an angular momentum quantum number $L$ for orbital excitations is added in parenthesis. So for instance
the $\psi(2D)$ is a $1^{--}$ charmonium state with $n=2$ and $L=2$; the spin is $S=1$ in order to have 
$C = (-)^{L+S} = -$ (in spectroscopic notation, $n\,^{(2S+1)}L_J = 2\,^3D_1$). Notice that $1^{--}$ states with odd $L$ are not allowed, 
being in contrast with the $P = (-)^{L+1}$ rule.

Charmonium and bottomonium states with $J^{PC}$ other than $1^{--}$ share the same naming conventions, based on the $S$ and $L$ quantum numbers, 
with a subscript $q = c$, $b$ to distinguish them. These are summarized in Table~\ref{tab:conventions} and a more detailed 
compilation can be found in Ref.~\refcite{pdg}. An additional subscript index can be added to distinguish states with the same $L$, $S$ 
numbers but different $J$. For instance, the $\chi_{c1}(2P)$ is a radially excited $L=1$, $S=1$ charmonium with $J=1$,
$P = (-)^{L+1} = +$ and $C = (-)^{L+S} = +$. Quarkonium-like states whose quantum numbers escape this classification are considered \emph{exotica}.

\begin{table}[ph]
\tbl{Naming conventions and allowed quantum numbers for standard quarkonia. The corresponding $J^{PC}$ quantum numbers are shown in parenthesis. 
The only $L=2$ states observed so far are the $1^{--}$ states, for which $\psi$ and $\Upsilon$ are used, as detailed in the text.}
{\begin{tabular}{@{}cccc@{}} \toprule
& L = 0 & L = 1 & L = 2 \\ 
\colrule
S = 0 & $\eta_{q}$ ($0^{-+}$) & $h_{q}$ ($1^{+-}$) & $\eta_{q2}$ ($2^{-+}$) \\
S = 1 & $\psi$, $\Upsilon$ ($1^{--}$) & $\chi_{qJ}$ ($0^{++}$, $1^{++}$, $2^{++}$) & ($1^{--}$, $2^{--}$, $3^{--}$)\\ 
\botrule
\end{tabular} \label{tab:conventions}}
\end{table}


\end{document}